# Causal Inference with Double/Debiased Machine Learning for Evaluating the Health Effects of Multiple Mismeasured Pollutants

Gang Xu[a], Xin Zhou[a,b], Molin Wang[c,d,e], Boya Zhang[e,f], Wenhao Jiang[a,g], Francine Laden[c,e,f], Helen H. Suh[h], Adam A. Szpiro[i], Donna Spiegelman[a,b], and Zuoheng Wang[a,j]

[a]Department of Biostatistics, Yale School of Public Health, New Haven, CT; [b]Center for Methods in Implementation and Prevention Science, Yale School of Public Health, New Haven, CT; [c]Department of Epidemiology, Harvard T.H. Chan School of Public Health, Boston, MA; [d]Department of Biostatistics, Harvard T.H. Chan School of Public Health, Boston, MA; [e]Channing Division of Network Medicine, Harvard Medical School, Brigham and Women's Hospital, Boston, MA; [f]Department of Environmental Health, Harvard T.H. Chan School of Public Health, Boston, MA; [g]Department of Biostatistics and Epidemiology, School of Public Health and Health Sciences, University of Massachusetts, Amherst, MA; [h]Department of Civil and Environmental Engineering, Tufts University, Medford, MA; [i]Department of Biostatistics, University of Washington, Seattle, WA; [j]Department of Biomedical Informatics & Data Science, Yale School of Medicine, New Haven, CT

CONTACT

Zuoheng Wang, Email: zuoheng.wang@yale.edu   Address: Department of Biostatistics, Yale University School of Public Health, New Haven, CT.

## ABSTRACT

One way to quantify exposure to air pollution and its constituents in epidemiologic studies is to use an individual's nearest monitor. This strategy results in potential inaccuracy in the actual personal exposure, introducing bias in estimating the health effects of air pollution and its constituents, especially when evaluating the causal effects of correlated multi-pollutant constituents measured with correlated error. This paper addresses estimation and inference for the causal effect of one constituent in the presence of other $PM_{2.5}$ constituents, accounting for measurement error and correlations. We used a linear regression calibration model, fitted with generalized estimating equations in an external validation study, and extended a double/debiased machine learning (DML) approach to correct for measurement error and estimate the effect of interest in the main study. We demonstrated that the DML estimator with regression calibration is consistent and derived its asymptotic variance. Simulations showed that the proposed estimator reduced bias and attained nominal coverage probability across most simulation settings. We applied this method to assess the causal effects of $PM_{2.5}$ constituents on cognitive function in the Nurses' Health Study and identified two $PM_{2.5}$ constituents, Br and Mn, that showed a negative causal effect on cognitive function after measurement error correction.

# 1. Introduction

The impact of air pollution, particularly particulate matter ≤ 2.5 μm ($PM_{2.5}$), on human health is substantial and well-documented (Garcia et al., 2023; Sukuman et al., 2023). Exposure to $PM_{2.5}$ has been associated with a range of adverse health effects affecting respiratory, cardiovascular, nervous, and other physiological systems. Recent studies reported that variations in the chemical composition of $PM_{2.5}$ play a role in modifying the association between $PM_{2.5}$ and health outcomes (Bell, 2012; Kim et al., 2022; Masselot et al., 2022; Zanobetti et al., 2009). Thus, it is important to identify the specific constituents of $PM_{2.5}$ that pose high health risks. However, it is difficult to evaluate the causal effects of individual $PM_{2.5}$ constituents with health, because humans are rarely, if ever, exposed to a single pollutant (Braun et al., 2016), and $PM_{2.5}$ constituents are often highly correlated and thus confound one another (Bell, 2012; Masselot et al., 2022). A comprehensive analysis that considers all $PM_{2.5}$ constituents simultaneously can help to eliminate potential confounding of each constituent by the others, offering more accurate causal interpretations of the effects of individual exposures. In addition, although $PM_{2.5}$ exposure measurement is usually estimated from spatio-temporal models (Kirwa et al., 2021; Yanosky et al., 2014), exposure measurements of $PM_{2.5}$ chemical constituents are often obtained from measurements made by the U.S. Environmental Protection Agency's (EPA) monitors nearest to individuals' residence in space and time in epidemiologic studies (Kazemiparkouhi et al., 2022), introducing substantial errors with respect to the actual exposure received by individuals, heretofore referred to as personal exposure, the quantity of interest. Measurement error can lead to bias in estimated effects of air pollution on health outcomes (Carroll et al., 2006; Spiegelman, 2010), especially when attempting to assess the causal effects of correlated multi-pollutant constituents of air pollution (Billionnet, Sherrill and Annesi-Maesano, 2012; Greenland, 1980; Zeka and Schwartz, 2004). Constituents with no effect but less measurement error, when correlated with poorly measured constituents with effects, can falsely appear to be strong risk factors, while the true risk factors can falsely appear to have no effect.

This study is motivated by an investigation of the impact of exposure to $PM_{2.5}$ and its constituents on cognitive decline in the Nurses' Health Study (NHS), an ongoing cohort study of U.S. female registered nurses. In 1976, NHS enrolled 121,700 women aged 30-55 years who resided in 11 U.S. states. Once recruited, participants completed a mailed questionnaire asking detailed questions about demographic characteristics, lifestyles and history of chronic diseases. Follow-up questionnaires are mailed to participants every two years to update information on lifestyle and health status (Colditz, Manson and Hankinson, 1997). From 1995 to 2001, a subset of NHS participants aged 70 years or older with no history of stroke were invited to a telephone-based study of cognitive function and completed their first cognitive interview. Three follow-up cognitive assessments were administered at approximately two-year intervals between 1995-2008. Previous studies in this cohort found a statistically significant association between long-term exposure to $PM_{2.5}$ and cognitive decline (Weuve et al., 2012). Literature reviews showed a connection between air pollution, including $PM_{2.5}$, and adverse neurological outcomes, such as cognitive impairment and an increased risk of neurodegenerative diseases (Delgado-Saborit et al., 2021; Weuve et al., 2021). A recent study conducted among Chinese older adults reported the adverse effects of long-term exposure to a mixture of five $PM_{2.5}$ constituents on cognitive impairment, including black carbon and organic matter (Qi et al., 2024). It is of great interest to determine the effects of individual of $PM_{2.5}$ constituents, both for a better understanding of the biological impacts of air pollution on health and for regulatory purposes. However, the impact of correlated multi-pollutant constituents of $PM_{2.5}$ on cognitive decline has not been investigated in NHS.

To assess the causal effects of individual $PM_{2.5}$ constituents on cognitive function, it is essential that the analysis adjusts for correlated multi-pollutant constituents with correlated measurement error. Statistical methods have been developed for the analysis of multi-pollutant air pollution. Existing methods broadly fall into two categories: dimension reduction methods and variable selection methods. Dimension reduction methods, such as principal component analysis (Gent et al., 2009) and positive matrix factorization (Krall and Strickland, 2017), transform a large number of correlated $PM_{2.5}$ constituents into a smaller set of factors to then be associated with health outcomes. These methods are not able to estimate the effect for individual $PM_{2.5}$ constituents. Another category of methods relies on variable selection to select the best subset of exposure variables, such as the least absolute shrinkage and selection operator (LASSO) (Sun et al., 2013b) and elastic net regression (Lenters et al., 2016). This category of methods allows for direct interpretation of the estimated effects for individual $PM_{2.5}$ constituents. However, it is well recognized in machine learning that regularization can reduce variance but introduces bias, potentially affecting the estimation of the parameter of interest. To achieve optimal efficiency in estimation, double/debiased machine learning (DML) methods (Chernozhukov et al., 2018) have been developed, employing two procedures: (1) orthogonalization to correct bias on the low-dimensional parameter of interest due to regularization on high-dimensional nuisance parameters, and (2) data splitting and cross-fitting to mitigate bias arising from overfitting. However, DML methods are not applicable to mismeasured exposures.

There are several methods for correcting measurement errors in air pollution epidemiology, including regression calibration (Hart et al., 2015; Liao et al., 2018), simulation extrapolation (Alexeeff, Carroll and Coull, 2016), and the parametric and non-parametric bootstrap (Bergen et al., 2013; Keller et al., 2017; Szpiro and Paciorek, 2013). In the context of multiple pollutants with measurement error, one approach considered a two-stage model within a hierarchical framework and used partial regression coefficients to assess the effect of each exposure (Schwartz and Coull, 2003; Zeka and Schwartz, 2004). While this method demonstrated unbiased estimation in the presence of measurement error, especially with a sufficiently large sample size, it can only handle two mismeasured pollutants. Another study utilized maximum likelihood principal component analysis to address classical measurement error using truncated singular value decomposition, assuming, unrealistically, that measurement errors follow independent normal distributions (Wentzell, Andrews and Kowalski, 1997; Wentzell and Lohnes, 1999). A recent study used penalized splines for exposure prediction and applied a non-parametric bootstrap for measurement error correction using geographic covariates in two-pollutant models (Bergen et al., 2016). However, most of these methods are not suitable for the main study (MS)/validation study (VS) design, where the MS provides the primary dataset for analysis and the VS provides additional data for measurement error correction (Cai et al., 2023; Hart et al., 2015; Liao et al., 2018; Rosner, Spiegelman and Willett, 1990; Spiegelman, Rosner and Logan, 2000). Unlike spatio-temporal models that use smoothing to estimate ambient air pollution exposure at individual residences, which would be the gold standard if individuals never left their houses, the VS collects personal exposure measurements, representing the actual exposure people receive, whether at home or not (Zhang et al., 2024). The measurement error in spatio-temporal models is how bad the smoothing models capture the exact exposure at the geographic locations of the participants and is a model misspecification error. These models are validated by comparing the estimated exposures with the actual measurements from the nearest monitors. In contrast, the MS/VS design corrects for bias in estimating the effects of mismeasured exposures in the MS by leveraging personal exposure measurements from the VS. To the best of our knowledge, methods do not exist for estimating the causal effects of individual $PM_{2.5}$ constituents in multi-pollutant analysis within the MS/VS design.

In this paper, we develop a novel method for analyzing multi-pollutant air pollution data using a flexible machine learning framework, addressing bias due to both measurement error and confounding by other constituents as well as other covariates not measured with error. Our focus is on evaluating the impact of one constituent in the presence of other $PM_{2.5}$ constituents, where all constituents are susceptible to measurement error and potential correlations in their underlying true values and their errors. Although machine learning methods, such as DML (Chernozhukov et al., 2018), can reduce bias in causal parameter estimation due to regularization and overfitting when adjusting for confounders, they cannot be directly applied to mismeasured exposures. Here, we first use a linear regression calibration model (Carroll et al., 2006; Rosner et al., 1990), fitted with generalized estimating equations (GEE), to estimate the measurement error process in an external validation study (EVS). Then, we extend the DML approach to correct for bias due to measurement error and consistently estimate the effect of interest in the MS. While motivated by the analysis of multi-pollutant air pollution data, this method is fully general for causal inference in multi-factorial studies where both the exposure of interest and confounders are subject to measurement error, possibly correlated. We organize the rest of the paper as follows: Section 2 develops the proposed DML method with bias correction for mismeasured exposures and mismeasured confounders, Section 3 illustrates the finite sample properties of the method through simulation studies, and in Section 4, we apply the method to NHS to assess the effects of multi-pollutant constituents of $PM_{2.5}$ on cognitive function. We close in Section 5 with discussion and conclusions.

## 2. Methods

As motivated by the data at hand, we consider an MS/EVS design where the participants in the MS and the EVS do not overlap.

The MS comprises observations $\{Y_i, \mathbf{Z}_i, \mathbf{W}_i\}$, for $i = 1, \dots, N$. For individual $i$ in the MS, let $Y_i$ denote a continuous outcome variable, and let $\mathbf{Z}_i = \left(Z_{1,i}, \mathbf{Z}_{2,i}^T\right)^T$ denote the surrogate exposures of all $PM_{2.5}$ constituents, where $Z_{1,i}$ is the surrogate exposure of the $PM_{2.5}$ constituent of interest, and $\mathbf{Z}_{2,i} = \left(Z_{2,1,i}, \dots, Z_{2,p,i}\right)^T$ is the surrogate exposures of other $PM_{2.5}$ constituents, with a length of $p$. Surrogate exposures are typically measured at the neatest monitors. $\mathbf{W}_i = \left(W_{1,i}, \dots, W_{q,i}\right)^T$ is a $q$-dimensional vector of potential confounders measured without error, such as age and sex.

The EVS comprises observations $\{\mathbf{X}_i, \mathbf{Z}_i, \mathbf{W}_i\}$, for $i = N+1, \dots, N+n$. The sample sizes in the MS and the EVS are denoted by $N$ and $n$, respectively. Usually, $N \gg n$, given the higher cost associated with measuring $\mathbf{X}_i$. For individual $i$ in the EVS, in addition to $\mathbf{Z}_i$ and $\mathbf{W}_i$, $\mathbf{X}_i = \left(X_{1,i}, \mathbf{X}_{2,i}^T\right)^T$ is available, representing the true exposures of all $PM_{2.5}$ constituents. Here, $X_{1,i}$ denotes the true exposure of the constituent of interest, and $\boldsymbol{X}_{2,i} = (X_{2,1,i}, \ldots, X_{2,p,i})^T$ denotes the true exposures of the other PM$_{2.5}$ constituents. True exposure is defined as personal exposure to PM$_{2.5}$ constituents of ambient origin.

The structure of this section is as follows: Section 2.1 gives an overview of the DML method, assuming true exposures are observed, Section 2.2 extends the DML method to correct for exposure measurement error, and Section 2.3 presents the statistical properties of the proposed method.

### 2.1. Double/debiased Machine Learning

To assess the effect of the exposure of interest on the outcome in the presence of other correlated constituents and potential confounders, we assume the following partially linear regression models:

$$Y_i = X_{1,i}\beta + g^*(\boldsymbol{X}_{2,i}, \boldsymbol{W}_i) + \xi_i, \tag{1}$$

$$X_{1,i} = m^*(\boldsymbol{X}_{2,i}, \boldsymbol{W}_i) + \epsilon_i. \tag{2}$$

Here, Eq. (1) is the outcome regression model, where $\beta$ is the parameter of interest, and Eq. (2) models the dependence of the exposure of interest $X_{1,i}$ on the other constituents $\boldsymbol{X}_{2,i}$ and potential confounders $\boldsymbol{W}_i$. The confounding variables $\boldsymbol{X}_{2,i}$ and $\boldsymbol{W}_i$ influence the values of the exposure of interest $X_{1,i}$ through the function $m^*(\cdot)$ and the outcome variable through the function $g^*(\cdot)$. The error terms $\xi_i$ and $\epsilon_i$ are assumed to be independent with $E(\xi_i|X_{1,i}, \boldsymbol{X}_{2,i}, \boldsymbol{W}_i) = 0$ and $E(\epsilon_i|\boldsymbol{X}_{2,i}, \boldsymbol{W}_i) = 0$.

If the true exposures $\boldsymbol{X}_i = (X_{1,i}, \boldsymbol{X}_{2,i}^T)^T$ had been measured in the MS, we could directly apply the DML approach (Chernozhukov et al., 2018) for estimation and inference on the parameter of interest, $\beta$. The resulting estimator is approximately consistent and normally distributed. The algorithm given by Chernozhukov et al. (2018) is as follows:

(1) Randomly split the $N$ subjects in the MS into two parts: an estimation sample $I$ of size $N_E$ and a training sample $I^c$ of size $N_T = N - N_E$, with $N_E/N \approx 0.5$.

(2) Estimate $\widehat{g^*}(\cdot)$ and $\widehat{m^*}(\cdot)$ from the training sample $I^c$ using machine learning methods such as LASSO (Tibshirani, 1996), ridge regression (Hoerl and Kennard, 1970), and random forests (Breiman, 2001).

(3) Construct an estimator of $\beta$ from the estimation sample $I$ using orthogonalization:

$$\hat{\beta}^{(1)} = \left[\sum_{i \in I} X_{1,i}\left(X_{1,i} - \widehat{m^*}(\boldsymbol{X}_{2,i}, \boldsymbol{W}_i)\right)\right]^{-1} \sum_{i \in I} \left(X_{1,i} - \widehat{m^*}(\boldsymbol{X}_{2,i}, \boldsymbol{W}_i)\right)\left(Y_i - \widehat{g^*}(\boldsymbol{X}_{2,i}, \boldsymbol{W}_i)\right).$$

(4) Perform cross-fitting by swapping the roles of the estimation and training samples to obtain a second version of the estimator, $\hat{\beta}^{(2)}$. Averaging the two estimators yields the final estimator $\hat{\beta} = (\hat{\beta}^{(1)} + \hat{\beta}^{(2)})/2$.

In the DML algorithm above, variable selection is conducted in Step (2) to choose the optimal subset of exposure variables and confounders that predict the exposure of interest and the outcome variable separately by minimizing predefined objective functions, such as the mean squared error with additional regularization terms. Orthogonalization and cross-fitting in Steps (3) and (4) are used to reduce the bias induced by regularization and overfitting in effect estimation in machine learning methods (Chernozhukov et al., 2018).

The above algorithm can be generalized to $K$-fold cross-fitting to obtain an alternative DML estimator as follows: (1) divide the $N$ subjects in the MS into $K \geq 2$ almost equal-sized random samples $(I_k)_{k=1}^K$, (2) compute $\hat{\beta}^{(k)}$, $k = 1, \dots, K$, using sample $I_k$ as the estimation sample and the remaining $K - 1$ samples $I_k^c$ as the training sample, and (3) compute the cross-fit estimator $\hat{\beta} = \frac{1}{K}\sum_{k=1}^K \hat{\beta}^{(k)}$. When the study sample size is small, it is recommended to choose a moderate value of $K$, such as 4 or 5, which provides more observations in the training sample $I_k^c$ to estimate the nuisance functions $g^*(\cdot)$ and $m^*(\cdot)$. Conversely, when the study sample size is large, as in our illustrative example where $N$ is 10,110, the choice of $K$ has minimal impact on finite sample performance (Chernozhukov et al., 2018). Consequently, we use $K = 2$ throughout this work.

### 2.2. Measurement Error Correction

In an MS/EVS design, the true exposures $\boldsymbol{X}_i = (X_{1,i}, \boldsymbol{X}_{2,i}^T)^T$ are not observed in the MS. Instead, the surrogate exposures $\boldsymbol{Z}_i = (Z_{1,i}, \boldsymbol{Z}_{2,i}^T)^T$ are available. Directly replacing the true exposures with the surrogate exposures in the model would introduce bias in estimation and inference due to measurement error (Carroll et al., 2006). One solution is to use empirical data on the measurement error process from the EVS to calibrate the exposure measurements in the MS. Following Carroll et al. (2006); Liao et al. (2018); Spiegelman et al. (2000), we assume transportability between the MS and the EVS, i.e., the distribution $f(\boldsymbol{X}_i|\boldsymbol{Z}_i, \boldsymbol{W}_i)$ for the measurement error process in the MS is the same as that in the EVS and the parameters of the measurement error model in the EVS can be reliably transported to the MS. Additionally, we assume that measurement error contains no extra information about the outcome beyond what is already provided by the true exposures, i.e., the outcome $Y_i$ is conditionally independent of the surrogate exposures $\boldsymbol{Z}_i$ given the true exposures $\boldsymbol{X}_i$ and the confounders $\boldsymbol{W}_i$. This is called the surrogacy assumption, or the nondifferential measurement error assumption (Boe et al., 2023; Carroll et al., 2006), which is usually satisfied when variables are measured prior to the outcome in a prospective cohort study (Carroll et al., 2006; Rosner et al., 1990). In our illustrative example, this holds since outcome data were not available when baseline exposure measurements were taken.

We assume a general regression calibration model for the measurement error process in the EVS, specified as

$$E(X_{1,i}|\boldsymbol{Z}_i, \boldsymbol{W}_i) = l(\boldsymbol{Z}_i, \boldsymbol{W}_i) \text{ and } E(\boldsymbol{X}_{2,i}|\boldsymbol{Z}_i, \boldsymbol{W}_i) = \boldsymbol{h}(\boldsymbol{Z}_i, \boldsymbol{W}_i), \qquad (3)$$

where $l(\cdot)$ is an unknown arbitrary scalar-valued function, and $\boldsymbol{h}(\cdot)$ is an unknown arbitrary $p$-dimensional vector-valued function. In practice, a linear regression calibration model is commonly considered (Cai et al., 2023; Hart et al., 2015; Kioumourtzoglou et al., 2014; Liao et al., 2018),

$$E(\boldsymbol{X}_i|\boldsymbol{Z}_i, \boldsymbol{W}_i) = (\boldsymbol{I}_{p+1} \otimes \boldsymbol{L}_i^T)\boldsymbol{\theta}, \qquad (4)$$

where $\boldsymbol{I}_{p+1}$ is the identity matrix of size $p + 1$, $\boldsymbol{L}_i = (1, \boldsymbol{Z}_i^T, \boldsymbol{W}_i^T)^T$ is a $(p + q + 2)$-dimensional vector consisting of the intercept, surrogate exposures for all constituents, and potential confounders, $\boldsymbol{\theta}$ is a vector of length $(p+1)(p+q+2)$ representing the parameters of the measurement error model, and $\otimes$ denotes the Kronecker product.

To estimate the parameter $\boldsymbol{\theta}$ in Eq. (4), we fit a GEE model,

$$\boldsymbol{U}(\boldsymbol{\theta}) = \sum_{i=N+1}^{N+n}(\mathbf{I}_{p+1} \otimes \boldsymbol{L}_i)\boldsymbol{V}_i^{-1}[\boldsymbol{X}_i - (\mathbf{I}_{p+1} \otimes \boldsymbol{L}_i^T)\boldsymbol{\theta}] = \mathbf{0}, \quad (5)$$

where $\boldsymbol{V}_i = \boldsymbol{A}^{1/2}\boldsymbol{R}(\alpha)\boldsymbol{A}^{1/2}$, with $\boldsymbol{A} = \text{diag}(\sigma_1^2, \ldots, \sigma_{p+1}^2)$ being the variances of true exposures of the $p + 1$ constituents, and $\boldsymbol{R}(\alpha)$ is a working correlation matrix that may depend on some unknown parameters $\alpha$. Note that $\boldsymbol{R}(\alpha)$ does not have to be correctly specified; for convenience, we use a working independence model with $\boldsymbol{R}(\alpha) = \mathbf{I}_{p+1}$ to avoid estimating many nuisance parameters. Since $\boldsymbol{V}_i$ are constant across subjects, we can show that under the working independence model, the values of variance parameters, $\sigma_1^2, \ldots, \sigma_{p+1}^2$, do not influence the estimation of $\boldsymbol{\theta}$. Equivalently, we then first estimate $\boldsymbol{\theta}$ by solving the estimating equations $\sum_{i=N+1}^{N+n}(\mathbf{I}_{p+1} \otimes \boldsymbol{L}_i)[\boldsymbol{X}_i - (\mathbf{I}_{p+1} \otimes \boldsymbol{L}_i^T)\boldsymbol{\theta}] = \mathbf{0}$. The corresponding estimator can be obtained as $\widehat{\boldsymbol{\theta}} = [\sum_{i=N+1}^{N+n}(\mathbf{I}_{p+1} \otimes \boldsymbol{L}_i\boldsymbol{L}_i^T)]^{-1}\sum_{i=N+1}^{N+n}(\mathbf{I}_{p+1} \otimes \boldsymbol{L}_i)\boldsymbol{X}_i$. Then, the variance parameters are estimated using the empirical sample variance estimators, given by

$$\hat{\sigma}_j^2 = \frac{1}{n}\sum_{i=N+1}^{N+n}\left[X_{i,j} - (\boldsymbol{e}_j \otimes \boldsymbol{L}_i)^T\widehat{\boldsymbol{\theta}}\right]^2, \quad \text{for } j = 1, \ldots, p+1,$$

where $\boldsymbol{e}_j$ is the vector of length $p + 1$ having $j$-th entry equal to 1 and all other entries 0.

The predicted values of the true exposures $\boldsymbol{X}_i$, denoted as $\widehat{\boldsymbol{X}}_i$, are then calculated as $\widehat{\boldsymbol{X}}_i = (\widehat{X}_{1,i}, \widehat{\boldsymbol{X}}_{2,i}^T)^T = (\mathbf{I}_{p+1} \otimes \boldsymbol{L}_i^T)\widehat{\boldsymbol{\theta}}$, for $i = 1, \ldots, N$ in the MS. More generally, $\widehat{\boldsymbol{X}}_i$ can be obtained by $\widehat{X}_{1,i} = \hat{l}(\boldsymbol{Z}_i, \boldsymbol{W}_i)$ and $\widehat{\boldsymbol{X}}_{2,i} = \widehat{\boldsymbol{h}}(\boldsymbol{Z}_i, \boldsymbol{W}_i)$ when the general regression calibration model of Eq. (3) is assumed.

Finally, we fit a modified partially linear regression models in the MS:

$$Y_i = \widehat{X}_{1,i}\beta + g(\widehat{\boldsymbol{X}}_{2,i}, \boldsymbol{W}_i) + \xi_i, \quad (6)$$

$$\widehat{X}_{1,i} = m(\widehat{\boldsymbol{X}}_{2,i}, \boldsymbol{W}_i) + \epsilon_i, \quad (7)$$

which differ from the models in Eq. (1) and (2) in Section 2.1 as follows: (1) the true exposures $X_{1,i}$ and $\boldsymbol{X}_{2,i}$ are replaced by the predicted values $\hat{X}_{1,i}$ and $\hat{\boldsymbol{X}}_{2,i}$; and (2) we use $g(\cdot)$ and $m(\cdot)$ instead of $g^*(\cdot)$ and $m^*(\cdot)$ to emphasize that they are generally different from one another. Under the transportability and surrogacy assumptions, it can be shown that the regression coefficients $\beta$ in Eq. (1) and (6) are equivalent when the following condition is satisfied (see details in Supplementary Materials S1):

$$E\big(g^*(\boldsymbol{X}_{2,i}, \boldsymbol{W}_i)\big|\boldsymbol{Z}_i, \boldsymbol{W}_i\big) = g(\boldsymbol{h}(\boldsymbol{Z}_i, \boldsymbol{W}_i), \boldsymbol{W}_i). \tag{8}$$

The estimate of the parameter of interest, $\beta$, is the solution to the score equation,

$$S(\hat{\boldsymbol{X}}; \beta, g, m) = \sum_{i=1}^{N} \big(\hat{X}_{1,i} - m(\hat{\boldsymbol{X}}_{2,i}, \boldsymbol{W}_i)\big)\big(Y_i - \hat{X}_{1,i}\beta - g(\hat{\boldsymbol{X}}_{2,i}, \boldsymbol{W}_i)\big) = 0. \tag{9}$$

A DML procedure, similar to that given in Section 2.1, can be used to obtain the DML estimator of $\beta$ from the estimation sample $I$, denoted as $\hat{\beta}_{RC}^{(1)}$, given by

$$\hat{\beta}_{RC}^{(1)} = \left[\sum_{i \in I} \hat{X}_{1,i}\big(\hat{X}_{1,i} - \hat{m}(\hat{\boldsymbol{X}}_{2,i}, \boldsymbol{W}_i)\big)\right]^{-1} \sum_{i \in I} \big(\hat{X}_{1,i} - \hat{m}(\hat{\boldsymbol{X}}_{2,i}, \boldsymbol{W}_i)\big)\big(Y_i - \hat{g}(\hat{\boldsymbol{X}}_{2,i}, \boldsymbol{W}_i)\big),$$

where $\hat{g}(\cdot)$ and $\hat{m}(\cdot)$ are the machine learning estimators of $g(\cdot)$ and $m(\cdot)$, respectively, from the training sample $I^c$. The final estimator is obtained by $\hat{\beta}_{RC} = \big(\hat{\beta}_{RC}^{(1)} + \hat{\beta}_{RC}^{(2)}\big)/2$, where $\hat{\beta}_{RC}^{(2)}$ is calculated as $\hat{\beta}_{RC}^{(1)}$ with the estimation and training samples reversed. Alternatively, we can perform $K$-fold cross-fitting to obtain a DML estimator of $\beta$.

### 2.3. Statistical Properties of $\hat{\boldsymbol{\beta}}_{RC}$

We focus on estimating $\beta$ in the partially linear regression model in Eq. (1). In the presence of measurement error, we establish theoretical results and derive the asymptotic variance for the DML estimator with regression calibration, $\hat{\beta}_{RC}$, in this section.

In an MS/EVS design, let $P$ be the probability law for the data $\boldsymbol{D} = \{Y, \boldsymbol{X}, \boldsymbol{Z}, \boldsymbol{W}\}$, where $\boldsymbol{X} = (X_1, \boldsymbol{X}_2^T)^T$ and $\boldsymbol{Z} = (Z_1, \boldsymbol{Z}_2^T)^T$, although the true exposure $\boldsymbol{X}$ is not observed in the MS. Under the transportability assumption, the probability law also applies to the data in the EVS, although the outcome $Y$ is not observed in the EVS. There are two sets of nuisance functions: $\boldsymbol{\gamma} = (l, \boldsymbol{h})$ and $\boldsymbol{\eta} = (g, m)$, where $\boldsymbol{\gamma}$ is the nuisance functions in the measurement error model estimated from the EVS, and $\boldsymbol{\eta}$ is the nuisance functions in the partially linear regression models estimated from the split-sample in the MS. Let $\beta_0$ denote the true value of $\beta$, and let $l_0, \boldsymbol{h}_0, g_0$ and $m_0$ denote the true values of the functions $l, \boldsymbol{h}, g$ and $m$ in Eq. (3), (6) and (7).

Let $(\delta_n)_{n=1}^\infty$ and $(\Delta_n)_{n=1}^\infty$ be sequences of positive constants approaching 0 such that $\delta_n \geq n^{-1/2}$. Also, let $c$ and $C$ be fixed strictly positive constants. In what follows, we use $\|\cdot\|_{P,r}$ to denote the $L^r(P)$ norm; that is, $\|f\|_{P,r} = \|f(\boldsymbol{D})\|_{P,r} = \left(\int |f(\boldsymbol{d})|^r dP(\boldsymbol{d})\right)^{1/r}$. In addition to the previously described transportability and surrogacy assumptions and Eq. (8), further assumptions are necessary to establish the consistency of $\hat{\beta}_{RC}$.

*Assumption 1.* For the probability law $P$ of $\boldsymbol{D} = \{Y, \boldsymbol{X}, \boldsymbol{Z}, \boldsymbol{W}\}$, the following conditions hold: (a) Models in Eq. (3), (6) and (7) hold; (b) the sample size ratio of the EVS to the MS, $n/N$, converges to a positive number $\lambda$ $(0 < \lambda < 1)$, as $N$ and $n$ converge to $\infty$; (c) $\|X_1\|_{P,r} \leq C$, $\|\xi\|_{P,r} \leq C$ and $\|\epsilon\|_{P,r} \leq C$, for $r = 2, 4$; (d) $|E(l_0 \cdot \epsilon)| \geq c$; and (e) given a random split-sample $I$ of size $N_E$, the nuisance parameter estimator $(\hat{\boldsymbol{\gamma}}, \hat{\boldsymbol{\eta}})$, obtained from the EVS and the training sample $I^c$, respectively, obeys the following conditions: With $P$-probability no less than $1 - \Delta_n$, 1) $\|\hat{l} - l_0\|_{P,2} \leq \delta_n$, 2) $\|\hat{l} - l_0\|_{P,4} \leq C$, 3) $\|\hat{g}(\hat{\boldsymbol{h}}, \cdot) - g_0(\boldsymbol{h}_0, \cdot)\|_{P,2} \leq \delta_n$, 4) $\|\hat{g}(\hat{\boldsymbol{h}}, \cdot) - g_0(\boldsymbol{h}_0, \cdot)\|_{P,4} \leq C$, 5) $\|\hat{m}(\hat{\boldsymbol{h}}, \cdot) - m_0(\boldsymbol{h}_0, \cdot)\|_{P,2} \leq \delta_n$, and 6) $\|\hat{m}(\hat{\boldsymbol{h}}, \cdot) - m_0(\boldsymbol{h}_0, \cdot)\|_{P,4} \leq C$.

Assumption 1(d) is essential for the proposed method. Noted in Eq. (7), $\hat{X}_{1,i} = \hat{l}(\boldsymbol{Z}_i, \boldsymbol{W}_i)$ and $\hat{\boldsymbol{X}}_{2,i} = \hat{\boldsymbol{h}}(\boldsymbol{Z}_i, \boldsymbol{W}_i)$ are both fixed functions of $\boldsymbol{Z}_i$ and $\boldsymbol{W}_i$. If $m(\cdot)$ is estimated from a rich set of functions, for example, random forest or deep neural network, the error term, $\epsilon_i$, may be very small or even zero, and Assumption 1(d) might be violated, which would lead to poor performance of the estimator $\hat{\beta}_{RC}$. So, we limit $m(\cdot)$ to linear functions in this work to satisfy this assumption. The estimates $\hat{g}(\cdot)$ and $\hat{m}(\cdot)$ are based on both the MS and the EVS. Since the sample size of the EVS is much smaller than that of the MS, it is the bottleneck for the rates of convergence. For Assumption 1(e), the bounds, $\delta_n$, for the rates of convergence hence depend on the sample size of the EVS, $n$. The rates of convergence of nuisance parameters are case specific, and for most machine learning methods, e.g. LASSO, regression trees, random forest, and deep neural networks, they are usually slower or equal to $1/\sqrt{n}$ (Belloni and Chernozhukov, 2013; Chernozhukov et al., 2024; Schmidt-Hieber, 2020; Wager and Athey, 2018; Wager and Walther, 2015), as implied by the construction of $\delta_n$.

The following theorem establishes the consistency of the DML estimator with regression calibration. The proof is provided in Supplementary Materials S2.

*Theorem 1.* When Assumption 1 holds, the DML estimator with regression calibration, $\hat{\beta}_{RC}$, constructed in Section 2.2 using the score in Eq. (9), converges in probability to $\beta_0$,

$$\hat{\beta}_{RC} - \beta_0 = O_P\left(N^{-\frac{1}{2}} + \delta_n\right).$$

To derive the asymptotic variance of $\hat{\beta}_{RC}$, besides the variability of the DML estimator when treating the predicted values $\hat{X}$ as fixed, it is essential to also account for the variability in the predicted values $\hat{X}$, which is calculated by fitting a regression calibration model to estimate $l(\cdot)$ and $\boldsymbol{h}(\cdot)$ from the EVS. Because the predicted values of the true exposures, $\hat{X}_i = \left(\hat{X}_{1,i}, \hat{X}_{2,i}^T\right)^T$, are calculated from nuisance functions, $\boldsymbol{\gamma} = (l, \boldsymbol{h})$, in the measurement error model estimated from the EVS, we rewrite the score function (9) as $S(\boldsymbol{D}; \beta, \boldsymbol{\gamma}, \boldsymbol{\eta}) = \sum_{i=1}^{N} S(\boldsymbol{D}_i; \beta, \boldsymbol{\gamma}, \boldsymbol{\eta})$, where $S(\boldsymbol{D}_i; \beta, \boldsymbol{\gamma}, \boldsymbol{\eta}) = \left(\hat{X}_{1,i} - m(\hat{\boldsymbol{X}}_{2,i}, \boldsymbol{W}_i)\right)\left(Y_i - \hat{X}_{1,i}\beta - g(\hat{\boldsymbol{X}}_{2,i}, \boldsymbol{W}_i)\right)$, to emphasize its dependence on $\boldsymbol{\gamma}$. As detailed in Supplementary Materials S3, the variance of $\hat{\beta}_{RC}$ is estimated as

$$\widehat{\mathrm{Var}}(\hat{\beta}_{RC}) \approx \left[\frac{1}{N} \sum_{k=1}^{2} \sum_{i \in I_k} \hat{X}_{1,i}\left(\hat{X}_{1,i} - \hat{m}^{(k)}(\hat{\boldsymbol{X}}_{2,i}, \boldsymbol{W}_i)\right)\right]^{-2}$$

$$\left\{\frac{1}{N^2}\sum_{k=1}^{2}\sum_{i\in I_k}\left[\left(\hat{X}_{1,i}-\hat{m}^{(k)}(\hat{X}_{2,i},W_i)\right)\left(Y_i-\hat{X}_{1,i}\hat{\beta}_{RC}-\hat{g}^{(k)}(\hat{X}_{2,i},W_i)\right)\right]^2+\right.$$

$$\left.\left[\frac{1}{N}\sum_{k=1}^{2}\sum_{i\in I_k}\frac{\partial S(D_i;\hat{\beta}_{RC},\gamma,\hat{\eta}^{(k)})}{\partial\gamma}\right]^T\widehat{\text{Var}}(\hat{\gamma})\left[\frac{1}{N}\sum_{k=1}^{2}\sum_{i\in I_k}\frac{\partial S(D_i;\hat{\beta}_{RC},\gamma,\hat{\eta}^{(k)})}{\partial\gamma}\right]\right\}\bigg|_{\gamma=\hat{\gamma}}, \quad (10)$$

by plugging in the estimated values $\hat{\beta}_{RC}$, $\hat{\gamma}$ and $\hat{\eta}^{(k)}$, and $\widehat{\text{Var}}(\hat{\gamma})$. The variance of $\hat{\beta}_{RC}$ in Eq. (10) consists of two components. The first component arises from the variance of the DML estimator $\hat{\beta}_{RC}$ when treating the predicted values $\hat{X}$ as fixed. Given the predicted values $\hat{X}$ in the MS, $\hat{\beta}_{RC}$ is asymptotically normal, as established by Theorem 4.1 in Chernozhukov et al. (2018). The second component arises from the variability induced by estimating $\gamma$ in the EVS.

In this study, we use GEE to fit a linear regression calibration model to estimate $l(\cdot)$ and $h(\cdot)$, and predict the true exposure $X$, as described in Section 2.2. The parameters in the linear regression calibration model (4) are $\theta$. Since $\hat{\theta}$ is the solution to the GEE of Eq. (5), we estimate $\text{Var}(\hat{\theta})$ using the sandwich variance estimator (Liang and Zeger, 1986) evaluated at the estimated value $\hat{\theta}$, given by

$$\widehat{\text{Var}}(\hat{\theta})\approx\left[-E\left(\frac{\partial U(\theta)}{\partial\theta}\right)\right]^{-1}\widehat{\text{Var}}(U(\theta))\left[-E\left(\frac{\partial U(\theta)}{\partial\theta}\right)\right]^{-1}\bigg|_{\theta=\hat{\theta}}$$

$$=\left[\sum_{i=N+1}^{N+n}\hat{V}_i^{-1}\otimes(L_iL_i^T)\right]^{-1}\left[\sum_{i=N+1}^{N+n}\left(\hat{V}_i^{-1}\widehat{\text{Var}}(X_i)\hat{V}_i^{-1}\right)\otimes(L_iL_i^T)\right]\left[\sum_{i=N+1}^{N+n}\hat{V}_i^{-1}\otimes\right.$$

$$\left.(L_iL_i^T)\right]^{-1}\bigg|_{\theta=\hat{\theta}}. \quad (11)$$

When fitting the working independence GEE model, $\hat{V}_i=\text{diag}(\hat{\sigma}_1^2,\ldots,\hat{\sigma}_{p+1}^2)$ is calculated using the empirical sample variance estimators, and $\text{Var}(X_i)$ is estimated by the sample covariance matrix, $\widehat{\text{Var}}(X_i)=[X_i-(I_{p+1}\otimes L_i^T)\hat{\theta}][X_i-(I_{p+1}\otimes L_i^T)\hat{\theta}]^T$. The variance of $\hat{\beta}_{RC}$ is then estimated by substituting $\theta$ in place of $\gamma$. Specifically, we replace $\widehat{\text{Var}}(\hat{\gamma})$ in Eq. (10) by $\widehat{\text{Var}}(\hat{\theta})$ from Eq. (11) and replace the derivatives with respect to $\gamma$ by those with respect to $\theta$. Since we estimate $\hat{\eta}^{(k)}=(\hat{g}^{(k)},\hat{m}^{(k)})$ using machine learning methods such as LASSO, $\frac{\partial S(D_i;\beta,\theta,\eta)}{\partial\theta}$ does not have a closed form. Instead, we use a numerical approach to approximate this derivative,

$$\frac{\partial S(\boldsymbol{D}_i; \beta, \boldsymbol{\theta}, \boldsymbol{\eta})}{\partial \theta_j} \approx \frac{S(\boldsymbol{D}_i; \beta, \theta_1, \ldots, \theta_{j-1}, \theta_j + \delta, \theta_{j+1}, \ldots, \theta_{(p+1)(p+q+2)}, \boldsymbol{\eta}) - S(\boldsymbol{D}_i; \beta, \boldsymbol{\theta}, \boldsymbol{\eta})}{\delta},$$

$$\text{for } i = 1, \ldots, N, \quad j = 1, \ldots, (p+1)(p+q+2),$$

where $\delta$ is set to 0.0001.

## 3. Simulation Studies

We conducted simulation studies to assess the finite sample performance of the proposed method under the MS/EVS design. Additionally, we compared this performance with that of an alternative estimator that did not used machine learning, to explore the potential advantages of machine learning in this context. Specifically, we compared two models: (1) DML with LASSO for variable selection, and (2) a saturated linear regression (SLR) model that includes all PM$_{2.5}$ constituents and covariates. For both models, we considered three fitting approaches: (i) true exposures (True), (ii) surrogate exposures ignoring measurement error (Uncorrected), and (iii) regression calibration for measurement error correction (Corrected). Altogether, we compared six methods: (1) DML-True; (2) uncorrected DML; (3) DML with regression calibration (DML-RC); (4) SLR-True; (5) uncorrected SLR; and (6) SLR with regression calibration (SLR-RC). In the simulations, we considered both linear and nonlinear effects of multi-pollutants on the outcome variable.

### 3.1. Simulation Settings

The sample sizes of the MS and the EVS were set to $N = 1,000$ and $n = 350$, respectively. Based on our illustrative example, in all simulations, we generated 12 PM$_{2.5}$ constituents ($X_{1,i}$ and $\boldsymbol{X}_{2,i}$), one continuous variable ($W_{1,i}$) representing age, one binary variable ($W_{2,i}$), and one continuous outcome variable ($Y_i$). In both datasets, the surrogate exposures $Z_{1,i}$ and $\boldsymbol{Z}_{2,i}$ were generated from the true exposures $X_{1,i}$ and $\boldsymbol{X}_{2,i}$ through the following model:

$$Z_{1,i} = X_{1,i} + \epsilon_i = X_{1,i,0} + 0.1 W_{2,i} + \epsilon_i, \quad Z_{2,j,i} = X_{2,j,i} + \tau_{j,i},$$

$$i = 1, \ldots, N + n; \quad j = 1, \ldots, 11,$$

where $X_{1,i,0}$ denotes the part of $X_{1,i}$ that is correlated with other PM$_{2.5}$ constituents, the positive random variables $(X_{1,i,0}, X_{2,i}^T) \sim MVN(\boldsymbol{\mu}, \boldsymbol{\Sigma})$ truncated at $\mathbf{0}$, $W_{1,i}$ is a continuous covariate generated from Unif$(18, 91)$ to mimic the range of age, $W_{2,i}$ is a binary confounding variable generated from a Bernoulli distribution with a probability of 0.16, and $(\epsilon_i, \tau_{1,i}, \ldots, \tau_{11,i}) \sim MVN(\mathbf{0}, \boldsymbol{\Sigma}_\epsilon)$ representing measurement error. The values of $\boldsymbol{\mu}, \boldsymbol{\Sigma}$, and $\boldsymbol{\Sigma}_\epsilon$ were set to the sample estimates from the Multi-Ethnic Study of Atherosclerosis (MESA) (Miller et al., 2019; Sun et al., 2013a), the EVS included in our illustrative example. We defined a minimum correlation value $\rho$, the parameter setting the extent of measurement error, with $\rho$ set at $\rho = 0.8, 0.6$, and $0.4$. We adjusted the variances of the error terms such that the correlations between the true exposure and the surrogate exposure of the 12 constituents were greater than $\rho$. These three scenarios represent low, moderate, and high amount of measurement error, respectively.

Continuous responses were generated in the MS under the outcome model:

$$Y_i = 8X_{1,i} + g^*(\boldsymbol{X}_{2,i}, \boldsymbol{W}_i) + \xi_i, \quad i = 1, \ldots, N,$$

where $\xi_i \sim N(0, \sigma_\xi^2)$ represents independent variation. We varied several aspects of the outcome model: (1) the function $g^*(\cdot)$ as linear and nonlinear, (2) the number of variables involved in $g^*(\cdot)$ as 1 and 4, and (3) $R^2$, the proportion of variance explained by predictors, took the values 0.4 and 0.6. In total, eight simulation scenarios were considered. In Scenarios 1-4, $g^*(\cdot)$ was a linear function, where $g^*(\boldsymbol{X}_{2,i}, \boldsymbol{W}_i) = 1 + 22X_{2,1,i}$ in Scenarios 1 and 2, and $g^*(\boldsymbol{X}_{2,i}, \boldsymbol{W}_i) = 1 + 12X_{2,1,i} + 12X_{2,2,i} + 12X_{2,3,i} + 12W_{2,i}$ in Scenarios 3 and 4. In Scenarios 5-8, $g^*(\cdot)$ was a nonlinear function following the logistic function in Chernozhukov et al. (2018), where $g^*(\boldsymbol{X}_{2,i}, \boldsymbol{W}_i) = 1 + 16 \frac{e^{20X_{2,1,i} - 0.4}}{1 + e^{20X_{2,1,i} - 0.4}}$ in Scenarios 5 and 6, and $g^*(\boldsymbol{X}_{2,i}, \boldsymbol{W}_i) = 1 + 8\frac{e^{20X_{2,1,i} - 0.4}}{1 + e^{20X_{2,1,i} - 0.4}} + 8\frac{e^{30X_{2,2,i} - 0.3}}{1 + e^{30X_{2,2,i} - 0.3}} + 8\frac{e^{500X_{2,3,i} - 0.02}}{1 + e^{500X_{2,3,i} - 0.02}} + 8W_{2,i}$ in Scenarios 7 and 8. The value of $\sigma_\xi^2$ was set at $\sigma_\xi^2 = 81$ in Scenarios 1 and 3, $\sigma_\xi^2 = 36$ in Scenarios 2 and 4, $\sigma_\xi^2 = 100$ in Scenarios 5 and 7, and $\sigma_\xi^2 = 49$ in Scenarios 6 and 8. All value combinations were set such that the average $R^2 \approx 0.4$ in Scenarios 1, 3, 5, and 7 and the average $R^2 \approx 0.6$ in Scenarios 2, 4, 6, and 8. For each scenario, 2,000 simulated datasets were generated. Table 1 summarizes the simulation scenarios, including the average $R^2$ and partial $R^2$.

**Table 1.** Description of Simulation Scenarios 1-8.

| Function $g^*(\cdot)$ | $g^*(X_{2,i}, W_i)$ | Simulation Scenario | # of Variables in $g^*(\cdot)$ | $R^2$ | Partial $R_X^2$ | Partial $R_g^2$ |
|---|---|---|---|---|---|---|
| Linear | $1 + 22X_{2,1,i}$ | S1 | 1 | 0.40 | 0.08 | 0.21 |
| | | S2 | 1 | 0.60 | 0.12 | 0.31 |
| | $1 + 12X_{2,1,i} + 12X_{2,2,i} + 12X_{2,3,i} + 12W_{2,i}$ | S3 | 4 | 0.40 | 0.08 | 0.23 |
| | | S4 | 4 | 0.60 | 0.12 | 0.35 |
| Nonlinear | $1 + 16 \frac{e^{20X_{2,1,i}-0.4}}{1+e^{20X_{2,1,i}-0.4}}$ | S5 | 1 | 0.42 | 0.06 | 0.26 |
| | | S6 | 1 | 0.59 | 0.09 | 0.36 |
| | $1 + 8\frac{e^{20X_{2,1,i}-0.4}}{1+e^{20X_{2,1,i}-0.4}} + 8\frac{e^{30X_{2,2,i}-0.3}}{1+e^{30X_{2,2,i}-0.3}} + 8\frac{e^{500X_{2,3,i}-0.02}}{1+e^{500X_{2,3,i}-0.02}} + 8W_{2,i}$ | S7 | 4 | 0.40 | 0.07 | 0.27 |
| | | S8 | 4 | 0.58 | 0.10 | 0.39 |

$R^2 = \frac{\text{Var}(X_1\beta + g^*(X_2, W))}{\text{Var}(Y)}$, partial $R_X^2 = \frac{\text{Var}(X_1\beta)}{\text{Var}(Y)}$, and partial $R_g^2 = \frac{\text{Var}(g^*(X_2, W))}{\text{Var}(Y)}$.

### 3.2. Simulation Results

To assess the finite sample performance of DML-RC for measurement error correction, we calculated the relative bias of $\hat{\beta}_{RC}$, defined as $(\bar{\hat{\beta}}_{RC} - \beta_0)/\beta_0$, where $\bar{\hat{\beta}}_{RC}$ was the average of $\hat{\beta}_{RC}$ over the 2,000 simulated datasets, and $\beta_0 = 8$ was the true value of $\beta$. The empirical standard error (ESE), the standard deviation of $\hat{\beta}_{RC}$ over the 2,000 simulated datasets, served as the benchmark for assessing the performance of the estimated asymptotic standard error (ASE) for $\hat{\beta}_{RC}$ derived in Eq. (11). Additionally, we calculated the empirical coverage probability of the 95% confidence interval containing the true value of $\beta$. We compared DML-RC to SLR-RC by comparing the bias ratio and the mean squared error (MSE) ratio between the two methods. To evaluate the bias resulting from exposure measurement error, uncorrected estimates were also calculated for both DML and SLR that ignore measurement error. Ideally, having access to the true exposures in the MS would provide upper bounds on the performance of DML and SLR. Therefore, we considered both models using the simulated true exposures to evaluate the best possible performance that could be expected when accurate exposure information is available.

The results for three levels of measurement error under the MS/EVS design are shown in Table 2 for Scenarios 1-4 and in Supplementary Table S1 for Scenarios 5-8. Notably, DML-RC had much smaller relative bias compared to the uncorrected DML. When true exposures were available, DML-True had the smallest relative bias among the three methods. The absolute value of the relative bias ranged from 0 to 0.003 for DML-True, 0.004 to 0.059 for DML-RC, and 0.226 to 0.684 for the uncorrected DML. The coverage probabilities of DML-True and DML-RC were close to 0.95, while the uncorrected DML had very low coverage probability, especially when the amount of measurement error was high or moderate. The ratio of ASE to ESE was close to 1 for DML-RC, indicating that the asymptotic variance of DML-RC approximated the ESE very well, in the presence of varying degrees of measurement error.

The absolute value of the bias ratio between DML and SLR was consistently below 1 when using regression calibration, suggesting that DML-RC effectively reduced bias while performing variable selection, outperforming SLR-RC. The coverage probability of DML-RC was close to 0.95, while SLR-RC had coverage probability below 0.90 in some simulation scenarios when the amount of measurement error was high or moderate. Additionally, the MSE ratio between DML and SLR was consistently below 1 when using regression calibration or true exposures. We found that both DML-RC and the uncorrected DML had smaller ASE compared to their SLR counterparts. However, the uncorrected DML showed slightly larger bias and MSE than the uncorrected SLR. These results demonstrated the advantages of combining DML and regression calibration for achieving more precise and less biased estimates in the presence of measurement error in the exposure and confounders.

**Table 2.** Simulation results for Scenarios 1-4 based on 2,000 simulation replicates.

| | $\rho$ | Input | SLR | | | DML | | | Bias ratio | MSE ratio |
|---|---|---|---|---|---|---|---|---|---|---|
| | | | RB | CP | ASE/ESE | RB | CP | ASE/ESE | | |
| S1 | | True | 0.000 | 0.951 | 1.002 | 0.000 | 0.952 | 0.998 | -0.124 | 0.760 |
| | 0.8 | Uncorrected | -0.209 | 0.434 | 0.999 | -0.228 | 0.276 | 0.976 | 1.091 | 1.126 |
| | | Corrected | -0.048 | 0.939 | 0.996 | -0.008 | 0.941 | 0.969 | 0.161 | 0.641 |
| | 0.6 | Uncorrected | -0.475 | 0 | 1.012 | -0.492 | 0 | 0.981 | 1.036 | 1.067 |
| | | Corrected | 0.131 | 0.888 | 1.012 | 0.047 | 0.920 | 0.969 | 0.360 | 0.503 |
| | 0.4 | Uncorrected | -0.660 | 0 | 1.005 | -0.680 | 0 | 0.976 | 1.030 | 1.059 |
| | | Corrected | 0.113 | 0.931 | 1.008 | 0.011 | 0.945 | 0.978 | 0.100 | 0.464 |
| S2 | | True | 0.000 | 0.951 | 1.002 | 0.000 | 0.952 | 1.001 | 0.140 | 0.759 |
| | 0.8 | Uncorrected | -0.208 | 0.144 | 0.999 | -0.228 | 0.051 | 0.976 | 1.093 | 1.160 |
| | | Corrected | -0.024 | 0.942 | 0.999 | 0.004 | 0.946 | 0.982 | -0.154 | 0.735 |
| | 0.6 | Uncorrected | -0.475 | 0 | 1.013 | -0.491 | 0 | 0.974 | 1.032 | 1.064 |
| | | Corrected | 0.131 | 0.828 | 1.013 | 0.046 | 0.912 | 0.971 | 0.348 | 0.404 |
| | 0.4 | Uncorrected | -0.660 | 0 | 0.997 | -0.678 | 0 | 0.961 | 1.027 | 1.054 |
| | | Corrected | 0.113 | 0.908 | 1.001 | 0.006 | 0.947 | 0.971 | 0.052 | 0.407 |
| S3 | | True | 0.003 | 0.956 | 1.018 | 0.002 | 0.944 | 0.979 | 0.726 | 0.819 |
| | 0.8 | Uncorrected | -0.228 | 0.352 | 1.019 | -0.231 | 0.253 | 0.988 | 1.013 | 1.001 |
| | | Corrected | 0.033 | 0.945 | 1.024 | 0.025 | 0.938 | 0.976 | 0.764 | 0.751 |
| | 0.6 | Uncorrected | -0.495 | 0 | 0.998 | -0.496 | 0 | 0.968 | 1.004 | 1.004 |
| | | Corrected | 0.075 | 0.924 | 1.005 | 0.016 | 0.935 | 0.942 | 0.208 | 0.633 |
| | 0.4 | Uncorrected | -0.677 | 0 | 0.953 | -0.684 | 0 | 0.932 | 1.009 | 1.018 |
| | | Corrected | 0.162 | 0.894 | 0.958 | 0.057 | 0.919 | 0.926 | 0.353 | 0.442 |
| S4 | | True | 0.002 | 0.956 | 1.018 | 0.001 | 0.945 | 0.981 | 0.502 | 0.819 |
| | 0.8 | Uncorrected | -0.229 | 0.072 | 1.022 | -0.232 | 0.040 | 0.988 | 1.015 | 1.017 |
| | | Corrected | 0.032 | 0.944 | 1.027 | 0.024 | 0.935 | 0.974 | 0.747 | 0.749 |
| | 0.6 | Uncorrected | -0.495 | 0 | 0.999 | -0.496 | 0 | 0.970 | 1.003 | 1.005 |
| | | Corrected | 0.074 | 0.907 | 1.005 | 0.012 | 0.940 | 0.942 | 0.167 | 0.559 |
| | 0.4 | Uncorrected | -0.678 | 0 | 0.955 | -0.684 | 0 | 0.929 | 1.008 | 1.016 |
| | | Corrected | 0.160 | 0.854 | 0.960 | 0.052 | 0.914 | 0.920 | 0.327 | 0.381 |

SLR: saturated linear regression; DML: doubly robust machine learning; RB: relative bias; CP: coverage probability; ASE: asymptotic standard error; ESE: empirical standard error; Bias ratio is the ratio of bias between DML and SLR; MSE ratio is the ratio of mean squared error between DML and SLR.

## 4. Analysis of the NHS Study

We applied our proposed method, DML-RC, to assess the causal effects of PM$_{2.5}$ constituents on cognitive function in NHS and compared these results to that of the uncorrected DML, SLR-RC, and the uncorrected SLR.

### *4.1. Data Description*

A subset of NHS participants underwent cognitive tests for general cognition, verbal memory, category fluency, working memory and attention, at up to four biennial assessments between 1995-2008. In this analysis, we included participants who completed the fourth cognitive assessment between 2004-2008. During the assessment, each participant completed a telephone interview of cognitive status (TICS) (Sterr et al., 2002) and five more cognitive tests, including the delayed recall of the TICS 10-word list, East Boston Memory Test (EBMT) -- immediate and delayed recalls, category fluency (animal naming test), and digit span backward (Baddeley et al., 1991). Due to the different scales among cognitive tests, for the outcome variable, we used z-score to create a global composite score as the average of all z-scores to evaluate the overall cognitive function.

We calculated the monthly average ambient exposure of $PM_{2.5}$ and 12 $PM_{2.5}$ constituents, Br, Ca, Cu, Fe, Mn, Ni, S, Se, Si, Ti, V and Zn, using concentrations measured at the monitors from the U.S. EPA database (Air Quality System, https://aqs.epa.gov/aqsweb/airdata/download_files.html) nearest to participants' residential addresses updated every two years over follow-up. From the monthly measurements, we calculated the 12-month moving average ambient exposure preceding the date of the fourth cognitive assessment for each participant and used it as the surrogate for the true exposure of $PM_{2.5}$ and its constituents. A detailed description of the study population, cognitive assessments, and exposure assessment in NHS has been previously reported (Weuve et al., 2012). After merging the surrogate exposure measurements of $PM_{2.5}$ and its constituents with the fourth cognitive assessment data in NHS, we retained 11,591 participants. We then excluded 1,082 participants due to missing surrogate exposure measurements for the 12 $PM_{2.5}$ constituents, resulting in a sample of 10,509 participants.

In the MESA VS, we assumed the true exposures are personal exposures to $PM_{2.5}$ and its constituents that were measured from 2005 to 2008 in nonsmoking households (Miller et al., 2019; Sun et al., 2013a). These exposures were measured over 5 consecutive days during both winter and summer across 1,092 days. All $PM_{2.5}$ samples were collected through active sampling on Teflon filters, and the concentrations of $PM_{2.5}$ constituents were determined using X-ray fluorescence. A total of 220 person-months were observed among 89 individuals. After the exclusion of 95 person-months due to missing personal exposure measurements for the 12 $PM_{2.5}$ constituents of interest and an additional 3 person-months due to missing surrogate exposure measurements for these constituents, we retained 122 person-months from 71 individuals, with paired measurements of personal and surrogate exposures available, serving as the EVS in our analysis.

*4.2. Data Processing and Analysis*

Following our previous work (Weuve et al., 2012), we considered several potential confounders, including age, body mass index (BMI), smoking status, and education, in both the MS and the EVS. In the MS, education was classified as having a registered nurse degree, bachelor's degree, and advanced degree; while in the EVS, education was categorized into multiple levels, ranging from no schooling (0) to graduate or professional school (8). To harmonize these different categorizations, we merged categories 0-6 (no schooling to an associate degree) into a single category in the EVS, comparable to the MS categorization. For missing values in BMI, smoking status, and education, mean imputation was performed for BMI and mode imputation for smoking status and education. Covariate missingness was handled by the missing covariate indicator method (Song et al., 2021).

We compared the surrogate exposures of total $PM_{2.5}$ and the 12 $PM_{2.5}$ constituents between the MS and the EVS and found that the mean and standard deviation (SD) of these exposures were comparable across the two studies, providing some empirical assurance for the transportability assumption. To stabilize the correlations between true and surrogate exposure for $PM_{2.5}$ and the 12 $PM_{2.5}$ constituents, we applied a log-transformation to both sets of measurements. Furthermore, we removed extreme outliers in both surrogate and personal exposure measurements. Extreme outliers were defined as values falling outside the quartiles $\pm 3 \times$ interquartile range (IQR). We treated outliers in surrogate and personal exposures differently to retain as much data as possible.

For surrogate exposures, we removed the entire observation of a participant from the analysis if one or more of their PM$_{2.5}$ constituents were extreme outliers. This led to the removal of 399 participants from the MS and 5 person-months from the EVS. For personal exposures, we removed only the specific extreme outlier measurement rather than the participant's entire data, since their remaining data could still be used for building the measurement error model for other constituents. This led to the removal of 33 personal exposure measurements from the EVS. Taken together, 10,110 participants in the MS and 117 person-months in the EVS were included in our analysis. Table 3 shows the basic characteristics of the final analysis samples in the MS and the EVS.

**Table 3.** Basic characteristics of the study population in the NHS fourth cognitive assessment (2004-2008) (Weuve et al., 2012) and MESA (Miller et al., 2019; Sun et al., 2013a).

| Characteristic | Main Study (NHS) $n = 10{,}110$ | External Validation Study (MESA) $n = 117$ |
|---|---|---|
| Age (years), mean (SD) | 80.74 (2.33) | 66.00 (8.18) |
| BMI (kg/m$^2$), mean (SD) | 25.42 (4.73) | 29.65 (5.07) |
| Smoking Status, $n$ (%) | | |
| – never | 4836 (48) | 54 (55) |
| – past | 4776 (47) | 38 (39) |
| – current | 462 (5) | 6 (6) |
| Highest Degree, $n$ (%) | | |
| – RN/AA | 7168 (77) | 58 (59) |
| – BA | 1477 (16) | 22 (23) |
| – MA/DR | 639 (7) | 18 (18) |

SD: standard deviation; RN: registered nurse; AA: Associate of Arts; BA: Bachelor of Arts; MA: Master of Arts; DR: doctorate.

We applied the proposed method to evaluate the causal effects of the 12 correlated constituents of PM$_{2.5}$ on cognitive function. In this analysis, each constituent was successively treated as the exposure of interest, adjusting for the confounding effects of total PM$_{2.5}$, the remaining PM$_{2.5}$ constituents, and the covariates. Two forms of confounding effects were considered in DML: (1) main effects only, and (2) main effects and all possible two-way interactions. To compare the performance of DML with SLR, we conducted a multi-pollutant analysis using SLR-RC, adjusting for the main effects of the confounders. Additionally, we performed a single-pollutant analysis using SLR-RC, where only the exposure of interest was considered, adjusting for total $PM_{2.5}$ and the potential confounding by age, BMI, smoking status, and education. Furthermore, uncorrected estimates were computed for both DML and SLR in both multi-pollutant and single-pollutant analyses, ignoring exposure measurement error.

### 4.3. Results

We examined the correlations among surrogate exposure measurements for total $PM_{2.5}$ and the 12 $PM_{2.5}$ constituents in the MESA VS (Supplementary Table S2). All $PM_{2.5}$ constituents showed positive correlations with total $PM_{2.5}$. Most $PM_{2.5}$ constituents were also positively correlated with each other. Measurement error, defined as the difference between log-transformed personal exposure and log-transformed surrogate exposure, was positively correlated between all $PM_{2.5}$ constituents and total $PM_{2.5}$ (Supplementary Table S3). The measurement errors of most $PM_{2.5}$ constituents were also positively correlated with each other. These correlations ranged from -0.16 to 0.73, with a median of 0.24. These findings confirm that $PM_{2.5}$ constituents were correlated, and their measurement errors were also correlated.

Table 4 reports the effect estimates, 95% confidence intervals, and p-values for the relationship between $PM_{2.5}$ constituents and cognitive function in NHS across the methods being compared. Significance was adjusted using the Benjamini-Hochberg false discovery rate (FDR) control to account for multiple hypothesis testing (Benjamini and Hochberg, 1995). $PM_{2.5}$ constituents with an FDR-adjusted p-value < 0.05 were considered statistically significant.

In the multi-pollutant analysis, Br exposure showed a statistically significant causal effect on cognitive function, but the direction of the effect changed after exposure measurement error correction. In the uncorrected analysis, we observed that increased Br exposure led to better cognitive function ($p = 0.001$, adjusted $p = 0.006$ for uncorrected DML with main effects). In contrast, the multi-pollutant analysis corrected for exposure measurement error found that increased Br exposure led to poorer cognitive function ($p = 0.000$, adjusted $p = 0.000$ for DML-

**Table 4.** Effect estimats, 95% confidence intervals, and p-values for the relationship between $PM_{2.5}$ constituents and cognitive function in NHS.

| Input | Uncorrected | | | | Corrected | | | |
|---|---|---|---|---|---|---|---|---|
| Method | SLR | | DML | | SLR | | DML | |
| | Single-pollutant | Multi-pollutant Main | Multi-pollutant Main | Multi-pollutant Interaction | Single-pollutant | Multi-pollutant Main | Multi-pollutant Main | Multi-pollutant Interaction |
| Br | 0.10 (0.04, 0.16) | 0.12 (0.04, 0.20) | 0.12 (0.04, 0.20) | 0.15 (0.03, 0.27) | 1.13 (0.54, 1.72) | -0.95 (-1.62, -0.28) | -1.00 (-1.53, -0.47) | -1.37 (-2.11, -0.63) |
| P-value | **0.000** | **0.002** | **0.001** | 0.016 | **0.000** | **0.005** | **0.000** | **0.000** |
| Ca | 0.01 (-0.01, 0.03) | 0.01 (-0.05, 0.07) | 0.01 (-0.05, 0.07) | 0.01 (-0.07, 0.09) | 0.03 (-0.03, 0.09) | 0.02 (-0.41, 0.45) | 0.03 (-0.42, 0.48) | -0.17 (-0.76, 0.42) |
| P-value | 0.668 | 0.809 | 0.740 | 0.713 | 0.230 | 0.911 | 0.883 | 0.575 |
| Cu | 0.00 (-0.02, 0.02) | -0.01 (-0.05, 0.03) | -0.01 (-0.03, 0.01) | 0.01 (-0.03, 0.05) | 0.05 (-0.05, 0.15) | 0.53 (0.06, 1.00) | 0.57 (0.08, 1.06) | 0.50 (-1.03, 2.03) |
| P-value | 0.909 | 0.517 | 0.652 | 0.561 | 0.384 | 0.027 | 0.021 | 0.517 |
| Fe | 0.01 (-0.01, 0.03) | 0.06 (0.00, 0.12) | 0.05 (-0.01, 0.11) | 0.05 (-0.05, 0.15) | 0.05 (-0.01, 0.11) | 0.11 (-0.09, 0.31) | 0.12 (-0.06, 0.30) | 0.10 (-0.14, 0.34) |
| P-value | 0.512 | 0.054 | 0.080 | 0.255 | 0.141 | 0.230 | 0.173 | 0.399 |
| Mn | 0.00 (-0.02, 0.02) | 0.00 (-0.04, 0.04) | 0.00 (-0.04, 0.04) | 0.00 (-0.06, 0.06) | 0.05 (-0.07, 0.17) | -0.66 (-1.15, -0.17) | -0.67 (-1.22, -0.12) | -0.64 (-1.05, -0.23) |
| P-value | 0.814 | 0.805 | 0.869 | 0.873 | 0.406 | **0.008** | 0.018 | **0.002** |
| Ni | -0.01 (-0.03, 0.01) | -0.02 (-0.04, 0.00) | -0.02 (-0.04, 0.00) | -0.02 (-0.06, 0.02) | -0.01 (-0.03, 0.01) | 0.22 (0.06, 0.38) | 0.24 (0.10, 0.38) | 0.24 (-0.09, 0.57) |
| P-value | 0.124 | 0.139 | 0.157 | 0.188 | 0.432 | **0.004** | **0.001** | 0.147 |
| S | 0.00 (-0.06, 0.06) | 0.02 (-0.04, 0.08) | 0.02 (-0.06, 0.10) | 0.05 (-0.05, 0.15) | -0.02 (-0.08, 0.04) | -0.06 (-0.28, 0.16) | -0.05 (-0.27, 0.17) | -0.10 (-0.37, 0.17) |
| P-value | 0.986 | 0.552 | 0.543 | 0.313 | 0.489 | 0.551 | 0.663 | 0.476 |
| Se | 0.02 (0.00, 0.04) | 0.00 (-0.04, 0.04) | 0.00 (-0.04, 0.04) | 0.07 (0.01, 0.13) | 0.01 (-0.01, 0.03) | -0.06 (-0.12, 0.00) | -0.06 (-0.12, 0.00) | -0.06 (-0.14, 0.02) |
| P-value | 0.098 | 0.884 | 0.845 | 0.032 | 0.253 | 0.065 | 0.046 | 0.203 |
| Si | 0.01 (-0.01, 0.03) | 0.00 (-0.04, 0.04) | 0.00 (-0.04, 0.04) | 0.03 (-0.03, 0.09) | 0.10 (-0.10, 0.30) | 0.17 (-0.12, 0.46) | 0.16 (-0.13, 0.45) | 0.34 (-0.17, 0.85) |
| P-value | 0.442 | 0.871 | 0.954 | 0.334 | 0.322 | 0.231 | 0.267 | 0.195 |
| Ti | -0.02 (-0.04, 0.00) | -0.07 (-0.11, -0.03) | -0.08 (-0.12, -0.04) | -0.08 (-0.16, 0.00) | -0.07 (-0.21, 0.07) | 0.14 (0.00, 0.28) | 0.14 (-0.02, 0.30) | 0.30 (0.08, 0.52) |
| P-value | 0.132 | **0.002** | **0.001** | 0.026 | 0.304 | 0.055 | 0.080 | **0.008** |
| V | -0.01 (-0.03, 0.01) | 0.02 (-0.02, 0.06) | 0.02 (-0.02, 0.06) | 0.02 (-0.04, 0.08) | 0.03 (-0.05, 0.11) | -0.14 (-0.30, 0.02) | -0.13 (-0.29, 0.03) | -0.12 (-0.32, 0.08) |
| P-value | 0.496 | 0.304 | 0.238 | 0.522 | 0.402 | 0.071 | 0.108 | 0.202 |
| Zn | 0.00 (-0.02, 0.02) | -0.04 (-0.08, 0.00) | -0.04 (-0.08, 0.00) | -0.04 (-0.10, 0.02) | 0.06 (-0.18, 0.30) | -0.07 (-0.27, 0.13) | -0.05 (-0.25, 0.15) | 0.01 (-0.19, 0.21) |
| P-value | 0.663 | 0.150 | 0.140 | 0.244 | 0.629 | 0.499 | 0.607 | 0.922 |

In multi-pollutant analysis, two forms of confounding effects were examined: (1) main effects only, denoted as "Multi-pollutant Main", and (2) main effects and two-way interaction effects, denoted as "Multi-pollutant Interaction". P-values are in bold if the corresponding FDR-adjusted p-values < 0.05.

RC). Similarly, the causal effect of Ti on cognitive function showed a shift in direction, from a negative effect in the uncorrected analysis to a positive effect after regression calibration. Additionally, Mn and Ni exposures showed no causal relationship with cognitive function in the uncorrected analysis. After correcting for exposure measurement error, we found that increased Mn exposure led to a decline in cognitive function ($p = 0.002$, adjusted $p = 0.012$ for DML-RC with interactions) and increased Ni exposure led to an improvement in cognitive function ($p = 0.001$, adjusted $p = 0.006$ for DML-RC with main effects). The effect sizes of these constituents in DML-RC were notably larger than those in the uncorrected DML. These results highlight the importance of correcting for exposure measurement error when assessing the causal effects of $PM_{2.5}$ constituents on cognitive function.

When comparing single-pollutant and multi-pollutant analyses, Br was the only constituent significantly associated with cognitive function in the single-pollutant analysis, regardless of exposure measurement error correction. In contrast, the multi-pollutant analysis identified additional $PM_{2.5}$ constituents, such as Mn, Ni, and Ti, associated with cognitive function. Furthermore, the effect of Br exposure on cognitive function reversed direction after adjusting for other $PM_{2.5}$ constituents and correcting for exposure measurement error.

In the uncorrected multi-pollutant analysis, SLR showed similar results to DML with main effects only. However, after correcting for exposure measurement error, increased Mn exposure was significantly correlated with worse cognitive function ($p = 0.008$, adjusted $p = 0.032$ for SLR-RC with main effects), while Mn showed no causal relationship with cognitive function in DML-RC with main effects. Mn had a negative causal relationship with cognitive function only when interactions were included in DML-RC. Additionally, when comparing DML-RC with main effects only to DML-RC with interactions, several $PM_{2.5}$ constituents, such as Mn, Ni, and Ti, showed a statistically significant causal effect on cognitive function in only one method. Br was the only constituent that showed a significant causal effect on cognitive function in both DML-RC methods. Because of the well-known dangers of overfitting, as in SLR, and of model misspecification, as in DML-RC with main effects only, the most reliable results are likely DML-RC with interactions.

## 5. Discussion

We investigated the influence of exposure to $PM_{2.5}$ constituents on cognitive function in NHS. Because $PM_{2.5}$ constituents are highly correlated with one another, we have confirmed that analysis of individual $PM_{2.5}$ constituents alone cannot accurately quantify its causal effects on health outcomes. To address this challenge, we developed a new method for analyzing multi-pollutant air pollution data using a machine learning framework. This approach corrects for exposure measurement errors on point and interval estimates, reduces bias due to regularization and overfitting, adjusts for confounding by a large number of covariates in a manner that avoids model misspecification, thereby also avoiding residual confounding, and evaluates the impact of correlated multi-pollutants on health outcomes. We used a linear regression calibration model, fitted with GEE in an EVS. Additionally, we extended the theory of the DML method originally developed by Chernozhukov et al. (2018) to correct for bias due to measurement error in the estimation of the effect of interest in the MS, establishing consistency of the new estimator. Furthermore, we demonstrated that the DML estimator with regression calibration is consistent and derived the variance estimator which included the uncertainty from both the outcome model and the measurement error model. Across nearly all the simulation settings considered, simulation results demonstrated that the new estimator had minimal bias, and its coverage probability was close to the nominal level. Applying this method to assess the impact of multi-pollutant constituents of $PM_{2.5}$ on cognitive function in NHS, our findings indicated that uncorrected analysis, ignoring measurement error, tended to underestimate the true impact of $PM_{2.5}$ constituents on cognitive function, potentially leading to spurious results. Conversely, in the multi-pollutant analysis with exposure measurement error correction, we identified several $PM_{2.5}$ constituents that statistically significantly influence cognitive function in NHS.

The correction for exposure measurement errors in our current multi-pollutant analysis of NHS was based on a relatively small EVS, which contains only 117 person-months. The limited sample size makes it difficult to obtain accurate estimates of the true exposures accounting for measurement error, potentially impacting the statistical power of our analysis and limiting the robustness and generalizability of our findings. To the best of our knowledge, the MESA VS is one of three validation studies of $PM_{2.5}$ constituents (Boomhower et al., 2022; Meng et al., 2005; Sun et al., 2013a) that exists in the world, with the other providing data too unreliable to use, and no additional validation studies are currently planned (Zhang et al., 2024). When it is possible to leverage larger EVS in future environmental studies, we anticipate an improved capacity to evaluate the causal effects of $PM_{2.5}$ constituents on health outcomes. The availability of a more extensive validation study will not only enhance the accuracy of exposure estimations but also contribute to a more comprehensive understanding of the complex relationship between air pollution and its impact on human health.

While motivated by the analysis of multi-pollutant air pollution data, this method is fully general and applicable to studies where both the exposures of interest and confounders are subject to measurement error and estimation of causal effects of individual exposures is the goal. The DML method with measurement error correction can efficiently be used to estimate the causal effect of the treatment or exposure of interest, adjusting for mismeasured confounders. The widespread prevalence of measurement error is a major threat to the reliability of epidemiology and public health research, often leading to biased results that can obscure true associations or causal effects. Our proposed method holds broad promise for causal inference in observational research in the presence of many confounders, most of which are measured with error, because it selects relevant confounders through efficient variable selection in a model robust manner, and corrects for measurement errors in both the exposure of interest and confounders, thereby enhancing the validity and efficiency of findings in observational causal inference research across a wide spectrum of health-related studies.

Our proposed DML-RC method has certain limitations. One limitation is its restriction to evaluate impacts on a single time point outcome, despite the NHS participants undergoing four cognitive assessments. Focusing solely on the fourth assessment in the current analysis may overlook valuable longitudinal information. A previous study from our group leveraged longitudinal $PM_{2.5}$ exposure to construct the exposure history and evaluated its impact on cognitive decline in NHS (Cai et al., 2023), demonstrating the potential value of exploring the longitudinal impact of air pollution on repeated cognitive function measures. Therefore, extending our methods to accommodate longitudinal multi-pollutant air pollution and health outcomes is an important direction for future research. Moreover, our current methodology is designed exclusively for continuous outcomes. The most important endpoints in NHS and other air pollution epidemiologic studies are all cause mortality, cardiovascular disease, and cancer. They are best analyzed as time-to-event outcomes. To address these challenges, we intend to extend our method to apply to survival outcome, providing a comprehensive toolkit for investigating environmental exposure effects on various outcomes in NHS, accounting for measurement error.

**Supplementary Materials**

The supplementary materials include the proof of Theorem 1, derivation of asymptotic variance, additional simulation results, and supporting results of data analysis. A Python package that implements DML-RC is provided at https://github.com/ZWang-Lab/DML-RC. The scripts used to conduct simulations presented in this article are available at https://github.com/ZWang-Lab/DML-RC_simulations.

**Funding**

This work was supported by the NIH under Grant R01ES026246. XZ was partially supported by NIH grant R03CA252808. ZW was partially supported by NIH grant R01LM014087-01S1.

**Disclosure Statement**

The authors report there are no competing interests to declare.

**Data Availability Statement**

The data that support the findings in this paper are not publicly available due to privacy or ethical restrictions.

**Table S1.** Simulation results for Scenarios 5-8 based on 2,000 simulation replicates.

| | $\rho$ | Input | SLR | | | DML | | | Bias ratio | MSE ratio |
|---|---|---|---|---|---|---|---|---|---|---|
| | | | RB | CP | ASE/ESE | RB | CP | ASE/ESE | | |
| S5 | | True | 0.000 | 0.951 | 1.002 | 0.000 | 0.951 | 0.998 | -0.071 | 0.760 |
| | 0.8 | Uncorrected | -0.214 | 0.488 | 0.998 | -0.226 | 0.352 | 0.975 | 1.055 | 1.058 |
| | | Corrected | 0.022 | 0.948 | 0.999 | 0.019 | 0.938 | 0.969 | 0.866 | 0.733 |
| | 0.6 | Uncorrected | -0.473 | 0 | 1.011 | -0.484 | 0 | 0.986 | 1.025 | 1.045 |
| | | Corrected | 0.120 | 0.913 | 1.011 | 0.044 | 0.928 | 0.970 | 0.368 | 0.544 |
| | 0.4 | Uncorrected | -0.654 | 0 | 1.007 | -0.671 | 0 | 0.984 | 1.026 | 1.051 |
| | | Corrected | 0.122 | 0.931 | 1.010 | 0.025 | 0.944 | 0.980 | 0.208 | 0.469 |
| S6 | | True | 0.000 | 0.949 | 1.021 | 0.000 | 0.952 | 0.998 | -0.155 | 0.794 |
| | 0.8 | Uncorrected | -0.207 | 0.305 | 1.018 | -0.228 | 0.162 | 0.983 | 1.104 | 1.170 |
| | | Corrected | -0.089 | 0.888 | 1.013 | -0.051 | 0.910 | 0.974 | 0.578 | 0.598 |
| | 0.6 | Uncorrected | -0.476 | 0 | 1.013 | -0.494 | 0 | 0.979 | 1.037 | 1.072 |
| | | Corrected | 0.137 | 0.852 | 1.014 | 0.051 | 0.914 | 0.969 | 0.370 | 0.454 |
| | 0.4 | Uncorrected | -0.663 | 0 | 1.000 | -0.683 | 0 | 0.970 | 1.030 | 1.059 |
| | | Corrected | 0.105 | 0.931 | 1.009 | 0.005 | 0.943 | 0.976 | 0.043 | 0.456 |
| S7 | | True | 0.003 | 0.956 | 1.018 | 0.003 | 0.943 | 0.976 | 0.900 | 0.823 |
| | 0.8 | Uncorrected | -0.226 | 0.434 | 1.018 | -0.226 | 0.358 | 0.987 | 0.999 | 0.972 |
| | | Corrected | 0.033 | 0.945 | 1.023 | 0.021 | 0.942 | 0.977 | 0.629 | 0.740 |
| | 0.6 | Uncorrected | -0.485 | 0 | 0.998 | -0.485 | 0 | 0.971 | 1.001 | 0.998 |
| | | Corrected | 0.081 | 0.928 | 1.005 | 0.024 | 0.933 | 0.942 | 0.296 | 0.639 |
| | 0.4 | Uncorrected | -0.664 | 0 | 0.954 | -0.670 | 0 | 0.936 | 1.010 | 1.018 |
| | | Corrected | 0.161 | 0.901 | 0.960 | 0.059 | 0.922 | 0.931 | 0.366 | 0.455 |
| S8 | | True | 0.002 | 0.956 | 1.018 | 0.002 | 0.944 | 0.979 | 0.761 | 0.821 |
| | 0.8 | Uncorrected | -0.227 | 0.154 | 1.021 | -0.227 | 0.088 | 0.989 | 1.000 | 0.987 |
| | | Corrected | 0.032 | 0.944 | 1.026 | 0.020 | 0.939 | 0.978 | 0.612 | 0.729 |
| | 0.6 | Uncorrected | -0.485 | 0 | 0.999 | -0.485 | 0 | 0.970 | 1.001 | 0.999 |
| | | Corrected | 0.081 | 0.908 | 1.005 | 0.021 | 0.935 | 0.944 | 0.265 | 0.573 |
| | 0.4 | Uncorrected | -0.664 | 0 | 0.955 | -0.670 | 0 | 0.933 | 1.009 | 1.017 |
| | | Corrected | 0.158 | 0.873 | 0.960 | 0.056 | 0.915 | 0.928 | 0.352 | 0.405 |

SLR: saturated linear regression; DML: doubly robust machine learning; RB: relative bias; CP: coverage probability; ASE: asymptotic standard error; ESE: empirical standard error; Bias ratio is the ratio of bias between DML and SLR; MSE ratio is the ratio of mean squared error between DML and SLR.

**Table S2.** Pearson correlation coefficients of log-transformed surrogate exposure of $PM_{2.5}$ and its constituents in the Multi-Ethnic Study of Atherosclerosis (MESA) validation study.

|        | $PM_{2.5}$ | Br    | Ca    | Cu    | Fe    | Mn    | Ni    | S     | Se    | Si    | Ti    | V     | Zn    |
|--------|------|-------|-------|-------|-------|-------|-------|-------|-------|-------|-------|-------|-------|
| $PM_{2.5}$ | 1.00 | 0.12  | 0.15  | 0.25  | 0.31  | 0.16  | 0.27  | 0.35  | 0.17  | 0.23  | 0.30  | 0.21  | 0.30  |
| Br     | 0.12 | 1.00  | 0.08  | 0.24  | 0.28  | 0.31  | 0.17  | 0.00  | 0.15  | -0.02 | 0.06  | 0.22  | 0.36  |
| Ca     | 0.15 | 0.08  | 1.00  | 0.69  | 0.74  | 0.76  | 0.48  | -0.28 | -0.40 | 0.73  | 0.61  | 0.62  | 0.56  |
| Cu     | 0.25 | 0.24  | 0.69  | 1.00  | 0.74  | 0.69  | 0.38  | -0.14 | -0.33 | 0.47  | 0.66  | 0.45  | 0.50  |
| Fe     | 0.31 | 0.28  | 0.74  | 0.74  | 1.00  | 0.73  | 0.48  | -0.03 | -0.28 | 0.61  | 0.67  | 0.56  | 0.69  |
| Mn     | 0.16 | 0.31  | 0.76  | 0.69  | 0.73  | 1.00  | 0.48  | -0.20 | -0.20 | 0.45  | 0.47  | 0.62  | 0.67  |
| Ni     | 0.27 | 0.17  | 0.48  | 0.38  | 0.48  | 0.48  | 1.00  | 0.11  | -0.19 | 0.20  | 0.30  | 0.68  | 0.60  |
| S      | 0.35 | 0.00  | -0.28 | -0.14 | -0.03 | -0.20 | 0.11  | 1.00  | 0.20  | 0.01  | 0.11  | -0.09 | -0.02 |
| Se     | 0.17 | 0.15  | -0.40 | -0.33 | -0.28 | -0.20 | -0.19 | 0.20  | 1.00  | -0.27 | -0.26 | -0.12 | -0.22 |
| Si     | 0.23 | -0.02 | 0.73  | 0.47  | 0.61  | 0.45  | 0.20  | 0.01  | -0.27 | 1.00  | 0.72  | 0.32  | 0.15  |
| Ti     | 0.30 | 0.06  | 0.61  | 0.66  | 0.67  | 0.47  | 0.30  | 0.11  | -0.26 | 0.72  | 1.00  | 0.32  | 0.28  |
| V      | 0.21 | 0.22  | 0.62  | 0.45  | 0.56  | 0.62  | 0.68  | -0.09 | -0.12 | 0.32  | 0.32  | 1.00  | 0.57  |
| Zn     | 0.30 | 0.36  | 0.56  | 0.50  | 0.69  | 0.67  | 0.60  | -0.02 | -0.22 | 0.15  | 0.28  | 0.57  | 1.00  |

**Table S3.** Pearson correlation coefficients of measurement error, defined as the difference between log-transformed personal exposure and log-transformed surrogate exposure, for PM$_{2.5}$ and its constituents in the Multi-Ethnic Study of Atherosclerosis (MESA) validation study.

|        | PM$_{2.5}$ | Br    | Ca    | Cu    | Fe   | Mn   | Ni    | S     | Se    | Si    | Ti    | V     | Zn    |
|--------|-----------|-------|-------|-------|------|------|-------|-------|-------|-------|-------|-------|-------|
| PM$_{2.5}$ | 1.00  | 0.34  | 0.41  | 0.21  | 0.32 | 0.06 | 0.10  | 0.40  | 0.08  | 0.24  | 0.27  | 0.17  | 0.20  |
| Br     | 0.34      | 1.00  | 0.25  | 0.13  | 0.37 | 0.17 | 0.25  | 0.51  | -0.07 | 0.19  | 0.21  | 0.06  | 0.27  |
| Ca     | 0.41      | 0.25  | 1.00  | 0.25  | 0.60 | 0.20 | 0.28  | 0.46  | -0.05 | 0.66  | 0.73  | 0.19  | 0.34  |
| Cu     | 0.21      | 0.13  | 0.25  | 1.00  | 0.42 | 0.12 | 0.11  | 0.23  | -0.10 | 0.37  | 0.28  | 0.27  | 0.41  |
| Fe     | 0.32      | 0.37  | 0.60  | 0.42  | 1.00 | 0.36 | 0.40  | 0.53  | 0.02  | 0.51  | 0.58  | 0.14  | 0.54  |
| Mn     | 0.06      | 0.17  | 0.20  | 0.12  | 0.36 | 1.00 | 0.19  | 0.26  | 0.02  | 0.16  | 0.15  | 0.17  | 0.14  |
| Ni     | 0.10      | 0.25  | 0.28  | 0.11  | 0.40 | 0.19 | 1.00  | 0.33  | -0.05 | 0.15  | 0.24  | 0.01  | 0.22  |
| S      | 0.40      | 0.51  | 0.46  | 0.23  | 0.53 | 0.26 | 0.33  | 1.00  | 0.16  | 0.36  | 0.25  | 0.23  | 0.36  |
| Se     | 0.08      | -0.07 | -0.05 | -0.10 | 0.02 | 0.02 | -0.05 | 0.16  | 1.00  | -0.16 | -0.15 | -0.08 | -0.15 |
| Si     | 0.24      | 0.19  | 0.66  | 0.37  | 0.51 | 0.16 | 0.15  | 0.36  | -0.16 | 1.00  | 0.69  | 0.24  | 0.32  |
| Ti     | 0.27      | 0.21  | 0.73  | 0.28  | 0.58 | 0.15 | 0.24  | 0.25  | -0.15 | 0.69  | 1.00  | 0.21  | 0.26  |
| V      | 0.17      | 0.06  | 0.19  | 0.27  | 0.14 | 0.17 | 0.01  | 0.23  | -0.08 | 0.24  | 0.21  | 1.00  | 0.10  |
| Zn     | 0.20      | 0.27  | 0.34  | 0.41  | 0.54 | 0.14 | 0.22  | 0.36  | -0.15 | 0.32  | 0.26  | 0.10  | 1.00  |